\documentclass[a4paper,fleqn]{cas-sc}

\usepackage[authoryear,longnamesfirst]{natbib}
\usepackage{siunitx}
\usepackage{longtable,tabularx}
\usepackage{bm}
\usepackage{subfig}
\usepackage{tikz}
\usetikzlibrary{arrows.meta,positioning,fit,backgrounds,calc}
\usepackage{algorithm}
\usepackage{algpseudocode}
\usepackage{lineno}

\begin{document}
\let\WriteBookmarks\relax
\def\floatpagepagefraction{1}
\def\textpagefraction{.001}

\shorttitle{Adaptive Attitude Estimation Using Light Curve Glints}
\shortauthors{Yoshimura and Hanada}

\title[mode=title]{Adaptive Attitude Estimation for Multiple-Surface Object Using Light Curve Glints}

\author[1]{Yasuhiro Yoshimura}
\cormark[1]
\ead{y.yoshimura.a64@m.kyushu-u.ac.jp}

\author[1]{Toshiya Hanada}
\ead{}

\affiliation[1]{organization={Department of Aeronautics and Astronautics, Kyushu University},
  addressline={744 Motooka, Nishi-ku},
  city={Fukuoka},
  postcode={819-0395},
  country={Japan}}

\cortext[1]{Corresponding author}

\begin{abstract}
  Light curve inversion enables the estimation of orbit, attitude, optical properties, and shape of space objects. Because a light curve is the temporal evolution of a scalar apparent magnitude, the estimation problem can be ill-posed owing to the non-uniqueness of the attitude that reproduces a given light curve. The initial estimate in Kalman filtering can therefore be sensitive, and, depending on the object properties, observation geometry, and number of estimated parameters, an inaccurate initial estimate may lead to divergence of the filter.
  A previous study uses a sudden change of light curves, called glint, to constrain the range of attitude estimate. The current paper extends the attitude estimation method using glint for multiple-surface objects. Such objects have multiple attitudes to yield glint, and the attitude estimate is not uniquely determined.
  To address this issue, this paper employs the interacting multiple model (IMM) algorithm that runs multiple parallel filters with model interaction in the estimation sequence.
  Each filter assumes that the glint occurs on the corresponding surface. The mode probability is updated by the likelihood of each filter, determining the correctness of the hypotheses. Furthermore, the mixing step in the IMM allows interaction among the filters through a transition probability matrix, enabling adaptation to the time-varying glint source. Numerical simulations are conducted for a box satellite in a geosynchronous orbit. Monte Carlo trials with initial attitude errors of up to 80~deg show that the proposed method improves the convergence rate from 10\% for a single surface filter to 73\%, and the mixing step is shown to be essential, since the convergence rate drops to 40\% when it is removed.
\end{abstract}

\begin{keywords}
  Attitude Estimation \sep Interacting Multiple Model \sep Light Curves \sep Space Situational Awareness
\end{keywords}

\maketitle

\section{Introduction}
Space situational awareness (SSA) is essential for maintaining the safety and sustainability of space operations. With the increasing number of active satellites and space debris~\citep{liou2022highlights}, obtaining accurate information about the states of space objects, including shape, attitude, and surface properties, has become important. Attitude estimation is particularly important for operational satellite monitoring and for preparation of active debris removal missions.
In addition, attitude information enables precise orbit propagation under attitude-dependent perturbation forces such as solar radiation pressure and atmospheric drag.

Light curve inversion is a cost-effective approach for estimating states of space objects, such as the attitude, optical properties, and shape, using photometric observations from ground-based telescopes~\citep{clarkResidentSpaceObject2022,burtonLightCurveAttitude2024, isolettaAttitudeMotionClassification2025}. A light curve is the temporal evolution of the apparent magnitude, which is a scalar quantity at each epoch, and the inversion can be ill-posed because different states may reproduce nearly the same light curve. The present paper focuses on attitude estimation, assuming the shape and optical properties to be known. The accuracy of the attitude estimate can be sensitive to the initial estimate supplied to the filter. Depending on the object properties, observation geometry, and number of estimated parameters, an inaccurate initial estimate may lead to divergence of the estimation filter.
To mitigate this sensitivity, recent research has focused on obtaining a sufficiently accurate initial estimate of the attitude. A particle swarm optimizer is used to estimate the initial attitude and angular velocity~\citep{burtonLightCurveAttitude2024}. \citet{gagnonParticleSwarmOptimization2025} also use a multiplicative particle swarm optimization based on quaternion multiplication to estimate the initial attitude.

A variety of estimation and optimization methods have been applied to light curve inversion. Early work used extended and unscented Kalman filters (UKF) to estimate the attitude from light curves~\citep{wettererAttitudeEstimationLight2009}, and multiple-model filters have been used to jointly estimate the attitude, shape, and surface properties~\citep{linaresSpaceObjectShape2014, dianettiResidentSpaceObject2023}. Batch and direct-inversion schemes have been studied to cope with measurement noise and to analyze the observability of the inversion~\citep{fanDirectLightCurve2020, friedmanObservabilityLightCurve2022}. More recently, machine learning has been applied to classify objects and their motion from light curves~\citep{linaresSpaceObjectsClassification2020, isolettaAttitudeMotionClassification2025}, and light curve inversion has been demonstrated on real measurements of spinnig debris~\citep{kucharskiFullAttitudeState2021}.

A previous study~\citep{matsushitaConceptualStudyImproved2024} proposes an attitude estimation method that exploits glint, a rapid change in light intensity caused by specular reflection. When a glint occurs, the geometric relationship among the Sun, the observer, and the surface normal of a space object is constrained within a specific range~\citep{matsushita}, providing a clue for the attitude estimate. Specular glints occur only under specific Sun--object--observer geometries, and their observability depends on the orbital regime. For objects in geosynchronous orbit (GEO), the slowly varying observation geometry yields repeatable glint opportunities during an observation pass, whereas for low Earth orbit (LEO) objects the rapidly changing geometry restricts glints to brief and less frequent events. Specular flashes have nonetheless been observed for both operational satellites and rotating debris~\citep{maleyVisualAppearanceIridium2003, silhaApparentRotationProperties2018, zilkovaSpaceDebrisSpectroscopy2023}, and the duration of a glint has been used to bound the angular velocity of an object~\citep{hinksAngularVelocityBounds2016}. By using an interval unscented Kalman filter, which projects the sigma points onto the glint constraint boundary, the previous method achieves fast convergence of the attitude estimate even with a large initial error. This state projection indicates that the glint constraint is applicable to determine the initial attitude estimate.
However, the previous method assumes a flat plate as the target object, which has only a single reflecting surface. For realistic space objects with multiple surfaces, such as box-wing satellites consisting of a bus body and solar panels, glint can originate from any of the multiple surfaces. When a glint is observed, the surface responsible for the glint is unknown, and this ambiguity prevents the direct application of the single-surface glint constraint~\citep{matsushitaConceptualStudyImproved2024}.

To address this challenge, this paper proposes an adaptive attitude estimation method using the interacting multiple model (IMM) algorithm~\citep{blom_interacting_1988}. The IMM framework runs multiple estimation filters in parallel, where each filter assumes that the observed glint originates from a different surface. Although the multiple model adaptive estimation (MMAE)~\citep{dianettiResidentSpaceObject2023} also uses multiple models, the IMM incorporates a mixing step that allows interaction among the filters through a transition probability matrix. This interaction enables the IMM to adapt to the time-varying glint source, as the surface causing the glint changes over time with the rotational motion of the space object. The likelihood of each filter provides a measure of the correctness of each hypothesis, and the combined estimate is computed as a weighted average. This approach naturally handles the ambiguity in the glint source without requiring prior knowledge of which surface causes the glint.

The paper is organized as follows. Section 2 presents the formulations of the BRDF, light curves, and attitude kinematics and dynamics. Section 3 describes the IMM-based attitude estimation method. Section 4 presents numerical simulations for a box satellite in a geosynchronous orbit to verify the effectiveness of the proposed method. Section 5 summarizes the conclusions.

\section{Formulations}
\subsection{Light curves}
\begin{figure}[tb]
  \begin{center}
    \includegraphics[width=0.8\linewidth]{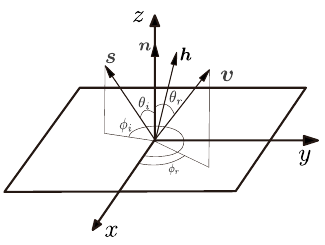}
  \end{center}
  \caption{Geometry of light reflection on a facet of a space object.}
  \label{fig:facet}
\end{figure}

A light curve is the time series of the apparent brightness of a space object observed by a ground-based telescope, and it is synthesized by modeling the sunlight reflected from each surface of the object toward the observer~\citep{wettererAttitudeEstimationLight2009, linaresSpaceObjectShape2014}. The reflectance of each surface is described by a bidirectional reflectance distribution function (BRDF).
A BRDF $f_{r}$ characterizes how a surface reflects incident light into a given reflection direction. It is defined as the ratio of the radiance $L_{r}$ reflected toward the observer to the irradiance $E_{i}$ incident on the surface,
\begin{align}
  f_r = \frac{L_{r}}{E_{i}}
\end{align}
which has units of $\mathrm{sr^{-1}}$.
The geometry of light reflection on a facet of a space object is illustrated in Fig.~\ref{fig:facet}, where the normal vector of the facet $\bm{n}$ is assumed to be along the $z$ axis. The Sun directional vector $\bm{s}$ and the observer directional vector $\bm{v}$ are unit vectors, defined by the sets of azimuth and polar angles ($\phi_{i},\theta_{i}$) and ($\phi_{r},\theta_{r}$), respectively.
The half vector between $\bm{s}$ and $\bm{v}$ is defined as
\begin{align}
  \bm{h} & = \frac{\bm{s}+\bm{v}}{\|\bm{s}+\bm{v}\|} \label{eq:bisector}
\end{align}
The azimuth and polar angles of $\bm{h}$ are also denoted as $\phi_{h}$ and $\theta_{h}$, respectively.

A BRDF contains two components, the diffuse term $c_{d}$ and the specular term $c_{s}$ as
\begin{align}
  f_{r} = c_{d} + c_{s}
\end{align}
The diffuse term gives a broad, nearly view-independent contribution to the light curve, whereas the specular term gives a sharp, strongly direction-dependent contribution that is responsible for glints.
This paper uses the Ashikhmin--Shirley model~\citep{ashikhminAnisotropicPhongBRDFc} for the BRDF, which is an anisotropic Phong model and is often used for light curves of space objects~\citep{gagnonParticleSwarmOptimization2025, jiangInversionSpaceDebris2023,haraAttitudeEstimationPhotometric2025}. It is a physically based microfacet model that satisfies energy conservation and Helmholtz reciprocity, and it assumes that each surface is an optically homogeneous flat facet with a possibly anisotropic specular lobe. The space object is represented as a collection of such flat facets;. The self-shadowing among facets is accounted for in the light curve computation, whereas secondary inter-facet reflections are neglected. The diffuse and specular terms are written as
\begin{align}
  c_{d} & = \frac{28\rho_{d}}{23 \pi} \left(1-F_{0}\right) \left[1-\left(1-\frac{\bm{n}^{T}\bm{s}}{2}\right)^{5}\right]\left[1-\left(1-\frac{\bm{n}^{T}\bm{v}}{2}\right)^{5}\right] \label{eq:Rd} \\
  c_{s} & = \frac{\sqrt{\left(n_{u}+1\right)\left(n_{v}+1\right)}}{8\pi} \frac{D(\bm{h}) F}{(\bm{v}^{T}\bm{h})\max(\bm{n}^{T}\bm{s},\bm{n}^{T}\bm{v})} \label{eq:Rs}
\end{align}
where $\rho_{d}\in[0,1]$ is the diffuse reflectance and is the fraction of incident light reflected diffusely. Also, $F_{0}$ is the Fresnel reflectance at normal incidence and $F$ is the Fresnel term, which increases the specular reflectance toward grazing angles. $D(\bm{h})$ is the microfacet normal distribution function, which determines the width and shape of the specular lobe.
These are written as
\begin{align}
  F         & = F_{0} + (1-F_{0}) (1 - \bm{v}^{T}\bm{h})^{5}                           \\
  D(\bm{h}) & = (\bm{n}^{T}\bm{h})^{n_{u}\cos^{2}{\phi_{h}} + n_{v}\sin^{2}{\phi_{h}}}
\end{align}
where $n_{u}$ and $n_{v}$ are the specular exponents that control the sharpness of the specular lobe along two orthogonal in-plane directions. Larger values yield a narrower, more mirror-like lobe, and $n_{u}=n_{v}$ corresponds to an isotropic specular lobe.

The reflected sunlight reaching the observer is obtained by summing the contributions of all visible facets~\citep{wettererAttitudeEstimationLight2009, linaresSpaceObjectShape2014}.
Given the direction of the Sun $\bm{s}$ and the normal vector of the $j$th facet $\bm{n}_{j}$, the power per unit square area of sunlight $E_{{\rm sun},j}$ reflected from the $j$th facet is described as
\begin{align}
  E_{{\rm sun},j} = C_{\rm sun}(\bm{n}_{j}^{T}\bm{s})f_{r,j}
\end{align}
where $C_{\rm sun}$ is the solar irradiance and $f_{r,j}$ is the BRDF of the $j$th facet. The power per unit square area of reflected sunlight received by the observer $E_{{\rm obs},j}$ is written as
\begin{align}
  E_{{\rm obs},j}= \frac{E_{{\rm sun},j}A_{j}(\bm{n}_{j}^{T}\bm{v})}{d^{2}}
\end{align}
where $A_{j}$ is the area of the $j$th facet and $d$ is the distance between the space object and the observer. Note that $E_{{\rm obs},j}$ has a nonzero value only when conditions
$\bm{n}_{j}^{T}\bm{s}>0$ and $\bm{n}_{j}^{T}\bm{v}>0$ are satisfied.
The apparent magnitude of the space object is
\begin{align}
  m_{\rm app} & = h(\bm{x}) + \varepsilon \label{eq:mapp}                                                                             \\
              & =m_{\rm sun} - 2.5\log_{10}\left|\sum_{j=1}^{N_{f}}\frac{E_{{\rm obs},j}}{C_{\rm sun}}\right| + \varepsilon \nonumber
\end{align}
where $h(\bm{x})$ is the observation function with the state vector $\bm{x}$, $\varepsilon$ is the observation noise assumed to follow a Gaussian distribution with zero mean and variance $R_{\rm true}$, $m_{\rm sun}=-26.7$ is the apparent magnitude of the Sun, and $N_{f}$ is the number of facets. The apparent magnitude is a logarithmic measure of the observed brightness of an object on the standard astronomical scale~\citep{bessellStandardPhotometricSystems2005, fankhauserSatelliteOpticalBrightness2023}, on which a smaller value corresponds to a brighter object. In this paper, the state vector $\bm{x}$ consists of the attitude only, as detailed in Section~\ref{sec:IMM}.

\subsection{Attitude kinematics and dynamics}\label{sec:attitudeDyn}
The rotational motion of a space object is defined with respect to the body-fixed frame relative to an inertial frame. The body-fixed frame has the origin at the object's center of mass and each axis corresponds to the principal axes of inertia. The $X, Y$, and $Z$ axes of the inertial frame are defined as follows: the $X$ axis is along the vernal equinox, the $Z$ axis points to the rotational axis of the Earth, and the $Y$ axis completes the right-handed system.

The kinematic equation of object attitude using quaternions is written as~\citep{markleyFundamentalsSpacecraftAttitude2014}
\begin{align}
  \dot{\bm{q}}
   & = \bm{f}(\bm{\omega},\bm{q}) \nonumber                                                     \\
   & =\frac{1}{2}\left[\begin{array}{cccc}
      0           & \omega_{z}  & -\omega_{y} & \omega_{x} \\
      -\omega_{z} & 0           & \omega_{x}  & \omega_{y} \\
      \omega_{y}  & -\omega_{x} & 0           & \omega_{z} \\
      -\omega_{x} & -\omega_{y} & -\omega_{z} & 0
    \end{array}\right]\left[\begin{array}{c}
      q_{1} \\
      q_{2} \\
      q_{3} \\
      q_{4}
    \end{array}\right]
\end{align}
where $\bm{q}=[q_{1},q_{2},q_{3},q_{4}]^{T}$ is the quaternion with scalar part $q_{4}$, $\bm{\omega}=[\omega_{x},\omega_{y},\omega_{z}]^{T}$ is the angular velocity of the body-fixed frame with respect to the inertial frame, and $\bm{f}(\bm{\omega},\bm{q})$ is the state propagation function.
The dynamics of rotational motion is described by Euler's equation as~\citep{markleyFundamentalsSpacecraftAttitude2014}
\begin{align}
  J\dot{\bm{\omega}} + \bm{\omega}\times (J\bm{\omega}) = \bm{\tau}
\end{align}
where $J$ is the inertia tensor and $\bm{\tau}$ is the disturbance torque in the body-fixed frame. For simplicity, the disturbance torques are ignored in this paper.
This torque-free assumption is appropriate for the short observation span considered in this paper. For an object in GEO, the dominant environmental torques are typically of order $10^{-6}$ to $10^{-4}~{\rm Nm}$~\citep{markleyFundamentalsSpacecraftAttitude2014, wertzSpacecraftAttitudeDetermination1978}, and the angular-velocity change they induce over a single pass is small relative to the initial rotational motion. The residual effect of these unmodeled torques is accounted for by the process noise $Q$ introduced in Section~\ref{sec:numericalSim}. Thus $Q$ is kept small in the present scenario and would need to be increased for longer observation spans or for objects subject to stronger perturbations, such as those with a high area-to-mass ratio, to preserve filter consistency.

\section{Method}
\subsection{Glint constraint for multiple facets}\label{sec:glintConstraint}
\begin{figure}[tb]
  \centering
  \includegraphics{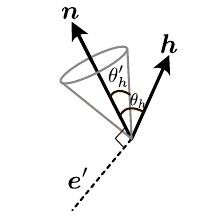}
  \caption{Projection of the estimated attitude onto the glint constraint boundary.}
  \label{fig:projection}
\end{figure}

A glint occurs when the specular component of reflected light becomes dominant. For a single facet with normal vector $\bm{n}_{j}$, the glint condition depends on the angle $\theta_{h,j}$ between $\bm{n}_{j}$ and the half vector $\bm{h}$~\citep{matsushita, matsushitaConceptualStudyImproved2024}. That is, a glint occurs when
\begin{align}
  \cos{\theta_{h}'} < \cos{\theta_{h,j}} = \bm{n}_{j}^{T}\bm{h} \label{eq:glintCondition}
\end{align}
where $\theta_{h}'$ is the threshold angle for glint occurrence, and it depends on the specular parameters $n_u$ and $n_v$ for the Ashikhmin--Shirley BRDF model. This constraint defines a range of attitude angles for which a glint can occur on the $j$th facet. Figure~\ref{fig:projection} illustrates the projection of the estimated attitude onto the glint constraint boundary, where $\bm{e}=\bm{h}_{b}\times\bm{n}_{j}$ is the rotation axis of projection and $\bm{h}_{b}$ is the half vector expressed in the body frame. The possible attitudes satisfying Eq.~\eqref{eq:glintCondition} are expressed as a cone centered on $\bm{n}_{j}$ with half-angle $\theta_{h}'$. Although there is still ambiguity in the projection direction, the attitude projection can reduce the estimation error. This projection is performed when the estimated attitude violates the glint constraint.

For a single-surface object, the glint constraint in Eq.~\eqref{eq:glintCondition} can be directly applied to project the attitude estimate~\citep{matsushitaConceptualStudyImproved2024} because the surface responsible for the glint is unambiguous. However, for a more complex object with $N_{f}$ facets, any of the $N_{f}$ facets may cause the glint. Thus, the surface responsible for the glint is not known a priori, and applying the wrong constraint would lead to incorrect estimation. This ambiguity necessitates a method that can simultaneously consider all possible glint sources.

\subsection{Unscented Kalman filter with glint constraint}
The attitude estimation uses the unscented Kalman filter (UKF)~\citep{julier} due to its applicability to nonlinear systems. The state variable is a quaternion, and the angular rate is assumed known. Modified Rodrigues parameters are used for calculating sigma points to satisfy the quaternion norm constraint~\citep{crassidis}. Throughout this paper, the estimated state is the attitude quaternion only. The angular velocity, the inertia tensor, the facet geometry, the BRDF optical parameters, and the orbital position are assumed known and held fixed during estimation.
When a glint is observed, the estimated attitude that violates the glint constraint in Eq.~\eqref{eq:glintCondition} can be projected onto the constraint boundary. Following the previous work~\citep{matsushitaConceptualStudyImproved2024}, the estimated state $\hat{{\bm{x}}}$ is projected as
\begin{align}
  \hat{\bm{x}} (k) \leftarrow \mathcal{P}\left[g_{j}\left(\hat{\bm{x}}(k)\right)\right] \label{eq:projection}
\end{align}
where $\mathcal{P}$ represents the projection operator, $k$ is the time instant, and $g_{j}(\hat{\bm{x}}(k))$ is the constraint function for the $j$th facet obtained from Eq.~\eqref{eq:glintCondition}:
\begin{align}
  g_{j}({\hat{\bm{x}}}) = \hat{\bm{n}}_{j}^{T}{\bm{h}} - \cos{\theta_{h}'} \label{eq:constraint}
\end{align}
where $\hat{\bm{n}}_{j}$ is the normal vector of the $j$th facet expressed in the inertial frame using the estimated attitude. Geometrically, the projection rotates the estimated quaternion minimally so that the angle $\theta_{h,j}$ between the half vector and the $j$th surface normal equals $\theta_h'$. When the estimated state already satisfies the constraint, i.e., $\theta_{h,j} \leq \theta_h'$, no projection is performed.

In addition to the mean state, the covariance is contracted along the projection direction to reflect the reduced attitude uncertainty implied by the glint constraint. Using the projection axis $\bm{e}=\bm{h}_{b}\times\bm{n}_{j}$ defined in Section~\ref{sec:glintConstraint}, the variance along this direction, $\sigma_{e}^{2}=\bm{e}^{T}P\bm{e}$, is limited to the squared angular tolerance $\theta_{h}'^{2}$ permitted by the constraint through the rank-one update
\begin{align}
  P \leftarrow P - \gamma\left(\sigma_{e}^{2}-\theta_{h}'^{2}\right)\bm{e}\bm{e}^{T} \qquad \sigma_{e}^{2}>\theta_{h}'^{2} \label{eq:covProjection}
\end{align}
where $0<\gamma\le 1$ controls the strength of the contraction. Directions orthogonal to $\bm{e}$ are left unchanged, so the constraint reduces the uncertainty only along the projection direction. This combined projection of the mean and covariance is consistent with constrained Kalman filtering, in which the estimate is projected onto the constraint surface and the covariance is reduced accordingly~\citep{simonKalmanFilteringState2002, simonKalmanFilteringState2010}. The choice of $\gamma$ in the IMM context is described in Section~\ref{sec:IMM}.

\subsection{Interacting multiple model (IMM) estimation}\label{sec:IMM}
\begin{figure}[tb]
  \centering
    \includegraphics{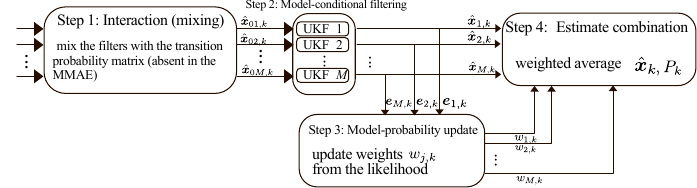}
  \caption{IMM attitude estimation sequence at each time step. }
  \label{fig:immBlock}
\end{figure}

The interacting multiple model (IMM) algorithm runs multiple estimation filters in parallel, each corresponding to a different hypothesis~\citep{blom_interacting_1988, zhang_detection_1998}. This paper uses the IMM to address the ambiguity in the glint source by assigning a dedicated filter to each surface hypothesis. Each filter assumes that the observed glint originates from a specific surface and projects the estimated state onto the corresponding glint constraint boundary. Although the MMAE algorithm also employs multiple filters, they operate independently of each other. In contrast, the IMM algorithm incorporates a mixing step that allows the filters to interact through a transition probability matrix. This interaction is essential because the surface responsible for the glint changes over time as the object rotates. Figure~\ref{fig:immBlock} shows the structure of the proposed estimator.

\begin{figure}[tb]
  \centering
  \includegraphics{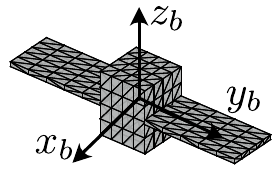}
  \caption{Box-wing satellite}
  \label{fig:bFrame}
\end{figure}

Dividing the object into small facets enables accurate calculation of light curves. However, many facets may have similar normal directions. For example, a box-wing satellite as shown in Fig.~\ref{fig:bFrame} can be grouped into six surfaces whose normal directions are along each axis of the body-fixed frame.
Let $M_{j}$ denote the hypothesis that the glint originates from the $j$th surface. The $j$th filter uses the constraint $g_{j}$ in Eq.~\eqref{eq:constraint} for the mean state projection. All filters share the same observation and the same state propagation model, but differ in the glint constraint applied during the observation update.

\subsubsection{IMM algorithm}
The IMM algorithm consists of the following four steps at each time step $k$, summarized in Algorithm~\ref{alg:imm}.
Step 1 is the mixing step, which computes a mixed initial condition for each filter using the transition probability matrix $P=[p_{ij}]$, where the components $p_{ij}$ represent the probability of switching from model $M_{i}$ to model $M_{j}$. The mixing probability $w_{ij,k}$ is computed as
\begin{align}
  w_{ij,k} = \frac{p_{ij}w_{i,k}}{c_{j,k}} \label{eq:mixprob}
\end{align}
where $w_{i,k}$ is the mode probability from the previous step. It is noted that if $p_{jj}=1$ for all $j$, the IMM reduces to the MMAE algorithm. The normalization constant $c_{j,k}$ is defined as
\begin{align}
  c_{j,k} = \sum_{i=1}^{M} w_{i,k}p_{ij} \label{eq:cbar}
\end{align}
The mixed initial state $\hat{\bm{x}}_{0j,k}$ and covariance $P_{0j,k}$ for the $j$th filter are computed as
\begin{align}
  \hat{\bm{x}}_{0j,k} & = \sum_{i=1}^{M} w_{ij,k} \hat{\bm{x}}_{i,k} \label{eq:mixstate}                                                                                         \\
  P_{0j,k}            & = \sum_{i=1}^{M} w_{ij,k} \left[P_{i,k} + (\hat{\bm{x}}_{i,k}-\hat{\bm{x}}_{0j,k}) (\hat{\bm{x}}_{i,k}-\hat{\bm{x}}_{0j,k})^{T}\right] \label{eq:mixcov}
\end{align}
This mixing step is the key difference from the MMAE and enables the filters to exchange information, preventing any single filter from diverging due to an incorrect hypothesis.

In Step 2, each filter performs state propagation and update of conventional UKF starting from the mixed initial condition $\hat{\bm{x}}_{0j,k} $ and covariance $P_{0j,k}$. When a glint is observed, the $j$th filter projects its estimated mean state using the constraint $g_{j}$ in Eq.~\eqref{eq:constraint}. When no glint is observed, each filter performs a measurement update of conventional UKF without projection. This step yields the filtered state $\hat{\bm{x}}_{j,k}$, the covariance $P_{j,k}$, and the innovation $\bm{e}_{j,k} = \bm{y}(k) - \hat{\bm{y}}_{j,k}^{-}$ for each filter.

In Step 3, the likelihood is computed from the innovation as
\begin{align}
  p(\bm{y}_k|\hat{\bm{x}}_{k,j}) = \frac{1}{\sqrt{\det(2\pi E_{j,k})}} \exp\left(-\frac{1}{2} \bm{e}_{j,k}^{T} E_{j,k}^{-1} \bm{e}_{j,k}\right) \label{eq:likelihood}
\end{align}
where $E_{j,k} = P_{yy,j,k}^{-} + R$ is the innovation covariance with the observation noise covariance $R$.
The mode probability $w_{j,k}$ is updated using the likelihood of each filter and the predicted mode probability $c_{j,k}$ in Eq.~\eqref{eq:cbar} as
\begin{align}
  w_{j,k} & = c_{j,k}\,p(\bm{y}_k|\hat{\bm{x}}_{k,j}) \label{eq:weight}              \\
  w_{j,k} & \leftarrow \frac{w_{j,k}}{\sum_{l=1}^{M} w_{l,k}} \label{eq:weight_norm}
\end{align}
where Eq.~\eqref{eq:weight_norm} means the normalization of the mode probabilities. The initial probability is set to a uniform distribution, i.e., $w_{j,0}=1/M$ for all filters, indicating no prior preference for any surface.

In Step 4, the combined state estimate $\hat{\bm{x}}_k$ and covariance $P_k$ are computed as a weighted average:
\begin{align}
  \hat{\bm{x}}_k & = \sum_{j=1}^{M} w_{j,k}\hat{\bm{x}}_{j,k} \label{eq:imm_state}                                                                               \\
  P_k            & = \sum_{j=1}^{M} w_{j,k} \left[P_{j,k} + (\hat{\bm{x}}_{j,k}-\hat{\bm{x}}_k)(\hat{\bm{x}}_{j,k}-\hat{\bm{x}}_k)^{T}\right] \label{eq:imm_cov}
\end{align}
These steps are repeated for each time step until the end of the estimation period.

\subsubsection{Transition probability matrix}
The transition probability matrix $P$ governs the interaction between the filters. In this study, the diagonal components are set to a large value $p_{jj}=p_{\rm same}$, reflecting the assumption that glint from the same surface is likely to persist over consecutive time steps. The off-diagonal components are set uniformly as $p_{ij}=(1-p_{\rm same})/(M-1)$ for $i\neq j$. This structure allows the IMM to smoothly transition between hypotheses as the glint source changes. Such a transition probability matrix, with a dominant self-transition probability and uniformly distributed switching probabilities, is a common choice in IMM design when no prior information favors a particular mode transition~\citep{barshalom, mazorInteractingMultipleModel1998}. The sensitivity of the estimation performance to $p_{\rm same}$ is examined in Section~\ref{sec:numericalSim}.

The mixing step provides the mixed initial conditions to each filter, and the $M$ parallel UKF filters perform updates. During a glint event, each filter projects its estimated mean state onto its respective constraint boundary. The mode probability update and estimate combination yield the final output, which also provides the mixed initial conditions to the next time step.
It is noted that the IMM does not require identification of the glint source surface. Instead, the likelihood mechanism automatically assigns higher mode probabilities to the filter with the correct hypothesis. The mixing step further enhances robustness by preventing filters with incorrect hypotheses from diverging, as they continuously receive corrected information from the other filters.

\begin{algorithm}[tb]
  \caption{IMM attitude estimation at time step $k$}
  \label{alg:imm}
  \begin{algorithmic}[1]
    \Require mode probabilities $w_{i,k}$, estimates $\hat{\bm{x}}_{i,k}$ and covariances $P_{i,k}$ ($i=1,\dots,M$), transition probability matrix $[p_{ij}]$, measurement $\bm{y}(k)$
    \Statex \emph{Step 1: Interaction (mixing)}
    \For{$j = 1$ to $M$}
      \State compute mixing probabilities $w_{ij,k}$ and $c_{j,k}$ \Comment{Eqs.~\eqref{eq:mixprob}, \eqref{eq:cbar}}
      \State form mixed initial condition $\hat{\bm{x}}_{0j,k},\,P_{0j,k}$ \Comment{Eqs.~\eqref{eq:mixstate}, \eqref{eq:mixcov}}
    \EndFor
    \Statex \emph{Step 2: Model-conditional filtering}
    \For{$j = 1$ to $M$}
      \State run the UKF prediction and update from $\hat{\bm{x}}_{0j,k},\,P_{0j,k}$
      \If{a glint is observed}
        \State project the mean and covariance onto the surface-$j$ constraint \Comment{Eqs.~\eqref{eq:projection}, \eqref{eq:covProjection}}
      \EndIf
      \State store $\hat{\bm{x}}_{j,k},\,P_{j,k}$ and the innovation $\bm{e}_{j,k}$
    \EndFor
    \Statex \emph{Step 3: Mode-probability update}
    \For{$j = 1$ to $M$}
      \State compute the likelihood $p(\bm{y}_k\,|\,\hat{\bm{x}}_{k,j})$ \Comment{Eq.~\eqref{eq:likelihood}}
      \State update $w_{j,k}=c_{j,k}\,p(\bm{y}_k\,|\,\hat{\bm{x}}_{k,j})$ \Comment{Eq.~\eqref{eq:weight}}
    \EndFor
    \State normalize the mode probabilities $w_{j,k}$ \Comment{Eq.~\eqref{eq:weight_norm}}
    \Statex \emph{Step 4: Estimate combination}
    \State combine $\hat{\bm{x}}_k,\,P_k$ as the weighted average \Comment{Eqs.~\eqref{eq:imm_state}, \eqref{eq:imm_cov}}
    \State \textbf{return} $\hat{\bm{x}}_k,\,P_k,\,w_{j,k}$
  \end{algorithmic}
\end{algorithm}

\section{Numerical Simulation} \label{sec:numericalSim}
\subsection{Simulation setup}
The target object is modeled as a box satellite with 1996 facets and 6 surfaces corresponding to the $\pm x_b$, $\pm y_b$, and $\pm z_b$ surfaces. The shape and surface parameters are summarized in Table~\ref{tb:ShapeParameters}. The specular lobe exponents are set to $(n_{u},n_{v})=(1000,1000)$ for all facets, representing isotropic specular reflectance.
These optical parameters are held fixed and assumed known during estimation. In practice, the optical properties of spacecraft materials degrade over time through space weathering, such as ultraviolet exposure, atomic-oxygen erosion, and surface contamination, which alters the diffuse reflectance $\rho_{d}$, the normal-incidence reflectance $F_{0}$, and the specular exponents~\citep{sharmaDegradationThermalControl2012, gotoChangesOpticalProperties2023}. A drift in these parameters would bias the synthesized light curve and therefore the attitude estimate. For long-term tracking, the filter would need to account for such changes. A quantitative analysis of the estimator's sensitivity to BRDF parameter uncertainty is left for future work.
\begin{table}[tb]
  \centering
  \caption{Shape and surface parameters.}
  \label{tb:ShapeParameters}
  \begin{tabular}{lcc}
    \hline\hline
    \multicolumn{3}{l}{\footnotesize\textit{Shape}} \\
    \footnotesize	Bus size                               &                                      & \footnotesize	 $1.0 \times 1.0 \times 1.0~{\rm m}$  \\
    \footnotesize	SAP size (each)                        &                                      & \footnotesize	 $5.0 \times 1.0 \times 0.02~{\rm m}$ \\
    \footnotesize	Total span (with SAPs)                 &                                      & \footnotesize	 $11.0~{\rm m}$ ($x_b$ direction)      \\
    \hline
    \multicolumn{3}{l}{\footnotesize\textit{Surface}} \\
    \footnotesize	Specular lobe exponents                & \footnotesize	$(n_{u},n_{v})$        & \footnotesize	$(1000,1000)$                   \\
    \footnotesize	Fresnel reflectance                    & \footnotesize	 $F_{0}$               & \footnotesize	 0.5                          \\
    \footnotesize	Diffuse reflectance                    & \footnotesize	 $\rho_{d}$            & \footnotesize	 0.5                          \\
    \hline
    \multicolumn{3}{l}{\footnotesize\textit{Inertia}} \\
    \footnotesize	Moment of inertia                      & \footnotesize	 $(J_{x},J_{y},J_{z})$ & \footnotesize	 $(450, 500, 800)~{\rm kgm^{2}}$ \\
    \hline\hline
  \end{tabular}
\end{table}
Table~\ref{tb:simCon} summarizes the simulation conditions. The true orbit is obtained by propagating a geosynchronous (GEO) satellite and its orbital elements are listed in Table~\ref{tb:simCon}. The true initial attitude and angular rate are set so that glint events occur on multiple surfaces during the simulation period. To verify the convergence capability of the proposed method under a poor initialization, the initial attitude estimate supplied to the filter is perturbed from the truth by an error whose magnitude is drawn from a uniform distribution $U[0,80]$~deg about a uniformly distributed random axis. The representative trials in Cases~1 and 2 use the same draw of about 56~deg from this distribution. The observation noise in the filter is inflated relative to the truth as $R = (3\sqrt{R_{\rm true}})^{2}$, and the system noise $Q$ is kept small, consistent with the torque-free dynamics assumed in Section~\ref{sec:attitudeDyn}.
The initial hypothesis probabilities are set uniformly as $w_{j,0} = 1/M$ for all filters.

The parameter values in Table~\ref{tb:simCon} are chosen as follows. The specular lobe exponents $(n_{u},n_{v})=(1000,1000)$ represent the near-mirror-like specular reflection of typical satellite surfaces and lie within the range used in previous light curve studies~\citep{gagnonParticleSwarmOptimization2025, jiangInversionSpaceDebris2023, haraAttitudeEstimationPhotometric2025}. The glint occurrence threshold $\theta_{h}'=7$~deg follows the value adopted in the previous paper~\citep{matsushitaConceptualStudyImproved2024}. The UKF scaling parameters $(\alpha,\beta,\kappa)=(10^{-3},2,0)$ are the standard values recommended for a Gaussian state distribution~\citep{julier, crassidis}. The small system noise $Q=10^{-12}$ reflects the assumed torque-free rotational dynamics, as discussed in Section~\ref{sec:attitudeDyn}. The transition probability $p_{\rm same}=0.99$ corresponds to a strong persistence of the glint source between consecutive steps. The estimation performance is shown to be insensitive to this value in Section~\ref{sec:numericalSim}.
The following examines two cases: in Case 1, most glints occur on the $+z$ surface, while in Case 2, glints occur on multiple surfaces. A sensitivity analysis of the transition probability $p_{\rm same}$ is then performed using 100 Monte Carlo runs.

\begin{table*}[tb]
  \centering
  \caption{Simulation conditions.}
  \label{tb:simCon}
  \begin{tabular}{lcc}
    \hline\hline
    \multicolumn{3}{l}{\footnotesize\textit{Orbit and observation}} \\
    \footnotesize	Semi-major axis              & \footnotesize	$a$                                   & \footnotesize	$4.2165\times10^{4}~{\rm km}$       \\
    \footnotesize	Eccentricity                 & \footnotesize	$e$                                   & \footnotesize	$2.16\times10^{-4}$                 \\
    \footnotesize	Inclination                  & \footnotesize	$i$                                   & \footnotesize	0.054 deg                           \\
    \footnotesize	Observer longitude, latitude & \footnotesize	$(\lambda_{\rm lon},\lambda_{\rm lat})$ & \footnotesize	(130.2165, 33.5946) deg            \\
    \footnotesize	Sampling interval            & \footnotesize	$\Delta t$                            & \footnotesize	5 s                                \\
    \hline
    \multicolumn{3}{l}{\footnotesize\textit{Initial attitude state (truth)}} \\
    \footnotesize	Initial attitude (3-2-1)     & \footnotesize	$(\psi,\theta,\phi)$                  & \footnotesize	$(90, -60, 30)$ deg                 \\
    \footnotesize	Initial quaternion           & \footnotesize	$\bm{q}_{0}$                          & \footnotesize	$[0.5, -0.1830, 0.6830, 0.5]^{T}$  \\
    \footnotesize	Initial angular rate         & \footnotesize	$\bm{\omega}_{0}$                     & \footnotesize	$[0.5, -0.5, 0.3]^{T}$ deg/s        \\
    \hline
    \multicolumn{3}{l}{\footnotesize\textit{Filter and scenario}} \\
    \footnotesize	Initial estimate error angle & \footnotesize	                                      & \footnotesize	$\sim U[0,80]$ deg (random axis)    \\
    \footnotesize	Glint occurrence threshold   & \footnotesize	$\theta_{h}'$                         & \footnotesize	7 deg                              \\
    \footnotesize	Number of surfaces           & \footnotesize	$M$                                   & \footnotesize	6                                  \\
    \footnotesize	UKF parameters               & \footnotesize	$(\alpha,\beta,\kappa)$               & \footnotesize	$(10^{-3}, 2, 0)$                  \\
    \footnotesize	Initial covariance           & \footnotesize	$P_{0}$                               & \footnotesize	$10^{-2}\bm{I}_{3\times 3}$         \\
    \footnotesize	Observation noise variance (truth) & \footnotesize	$R_{\rm true}$                  & \footnotesize	$0.1~{\rm mag^{2}}$                 \\
    \footnotesize	Filter observation noise     & \footnotesize	$R$                                   & \footnotesize	$(3\sqrt{R_{\rm true}})^{2}=0.9~{\rm mag^{2}}$ \\
    \footnotesize	System noise                 & \footnotesize	$Q$                                   & \footnotesize	$10^{-12}$                         \\
    \footnotesize	Transition probability       & \footnotesize	$p_{\rm same}$                        & \footnotesize	0.99                               \\
    \hline \hline
  \end{tabular}
\end{table*}

\subsection{Case 1: Most of the glints occur on the $+z$ surface}
\begin{figure}[tb]
  \centering
  \includegraphics[width=7.5cm]{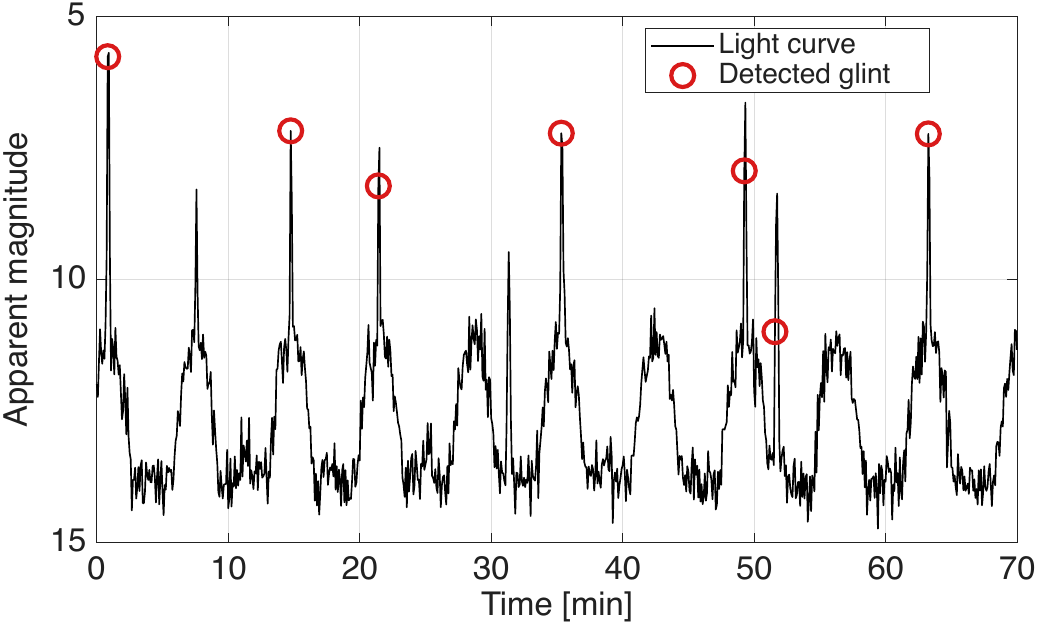}
  \caption{Light curve of the box satellite (Case 1).}
  \label{fig:lightcurveRep01}
\end{figure}

\begin{figure}[tb]
  \centering
  \includegraphics[width=7.5cm]{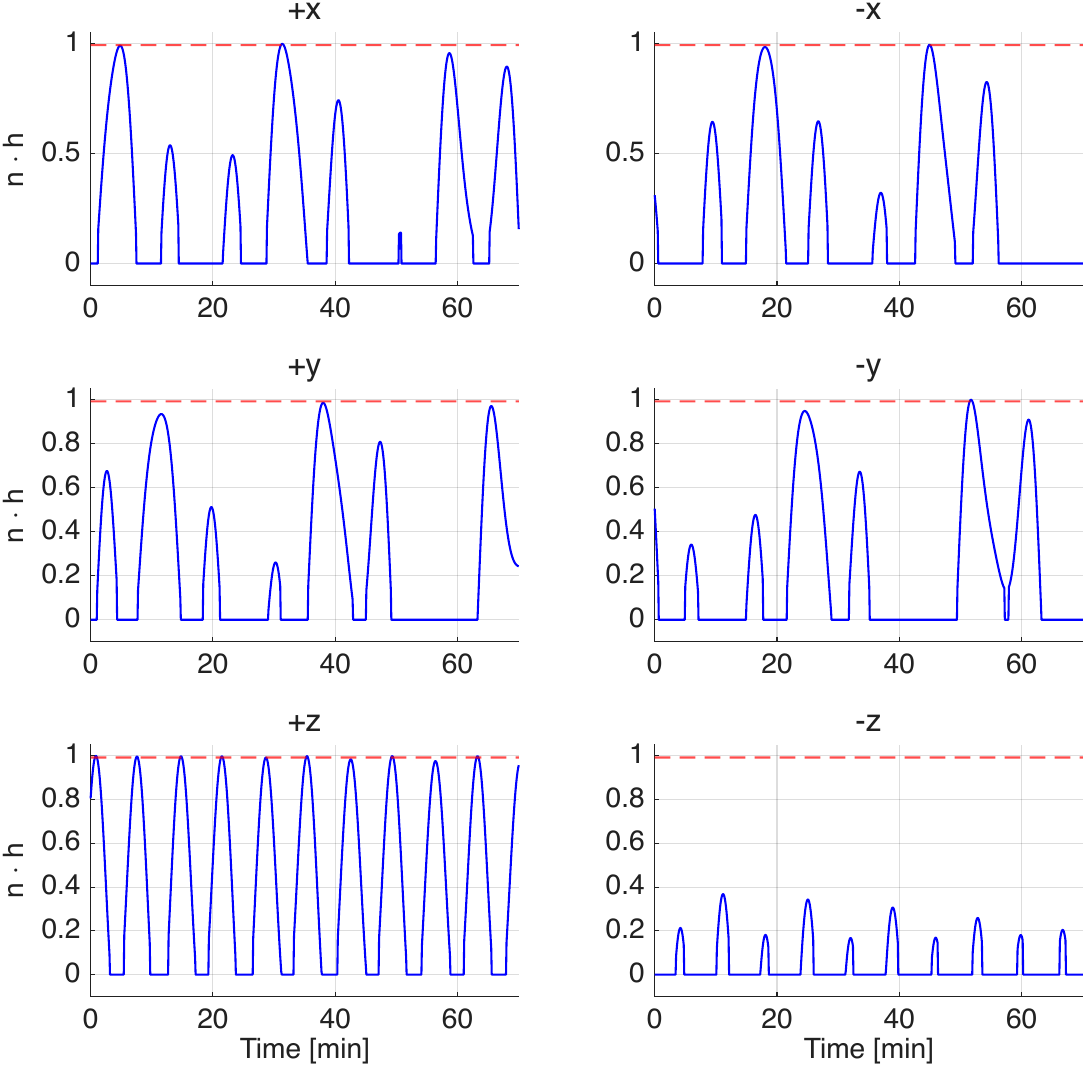}
  \caption{The inner product between each normal vector and the half vector (Case 1).}
  \label{fig:halfVecRep01}
\end{figure}

Figure~\ref{fig:lightcurveRep01} shows the simulated light curve of the box satellite. Several glint events appear as sharp peaks in the light curve, occurring when the half vector aligns with a surface normal. 
Figure~\ref{fig:halfVecRep01} shows the inner product between each normal vector and the half vector, and glints occur when its value becomes one. 
In this case, most of the glints occur on the $+z$ surface because the $+z$ panel in Fig.~\ref{fig:halfVecRep01} reaches one periodically.
Thus, even if the single surface UKF method is used, it may be able to estimate the attitude accurately.

\begin{figure}[tb]
  \centering
  \includegraphics[width=7.5cm]{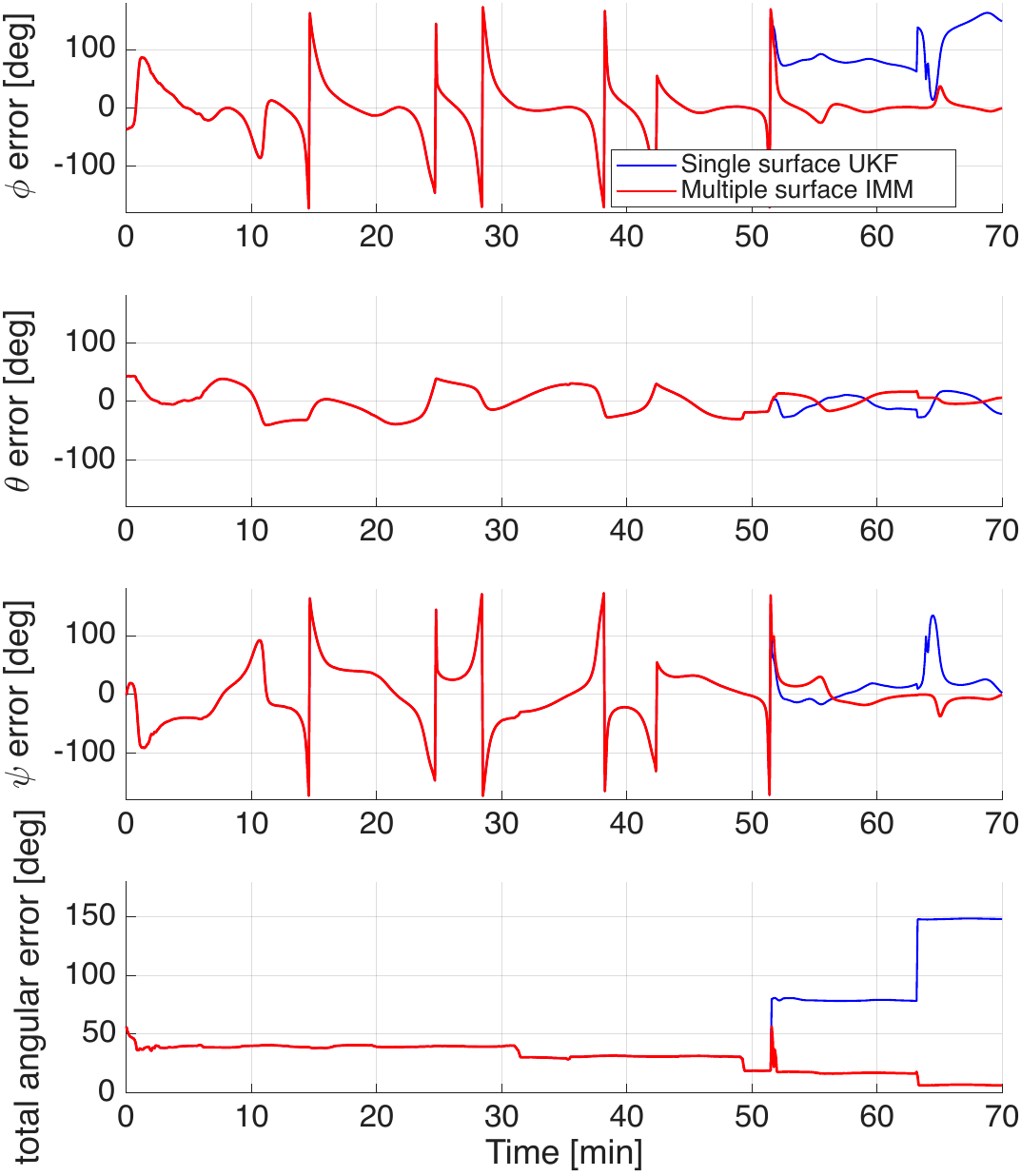}
  \caption{Euler angle error using multi surface IMM and single surface IMM (Case 1).}
  \label{fig:eulerErrRep01}
\end{figure}

\begin{figure}[tb]
  \centering
  \includegraphics[width=7.5cm]{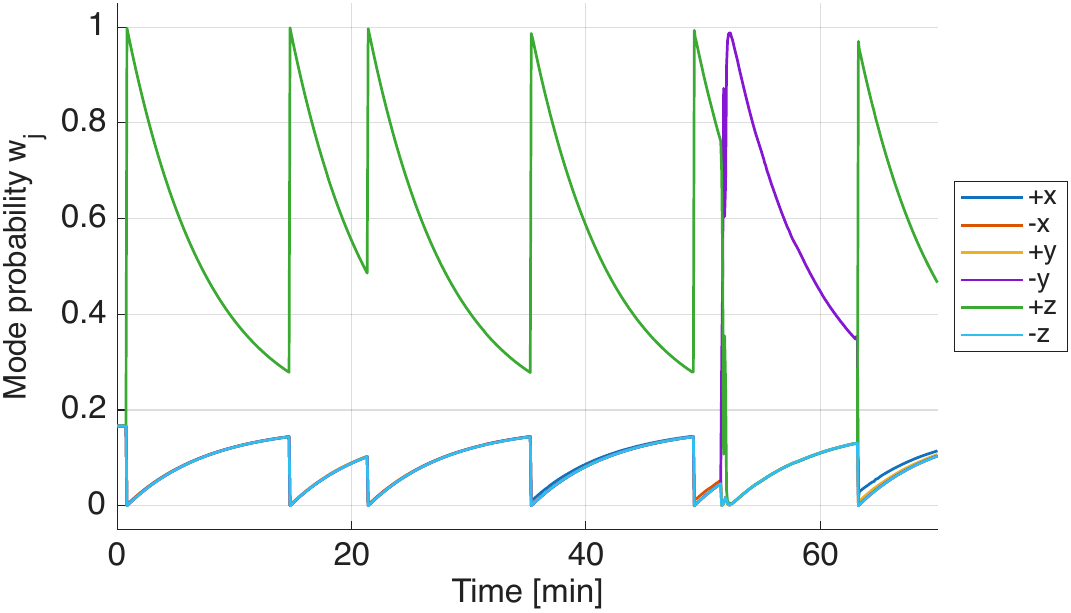}
  \caption{Time history of mode probabilities $w_{j}$ for the multiple surface IMM method (Case 1).}
  \label{fig:modeProbRep01}
\end{figure}

Figure~\ref{fig:eulerErrRep01} shows the attitude estimation errors for the previously proposed single surface UKF~\citep{matsushitaConceptualStudyImproved2024} and the proposed multi surface IMM method. The upper threee figures shows the error in the 3-2-1 Euler angle sequence $(\phi\rightarrow\theta\rightarrow\psi)$, and the bottom figure shows the total attitude error $\Phi$, i.e., the principal rotation angle of the error quaternion. The large $\pm180^\circ$ swings in the individual Euler angles arise from the wrapping near singular orientations and do not represent the actual attitude error, which is captured by $\Phi$. The detected glint timings are marked with red circles in Figure~\ref{fig:lightcurveRep01}.
Since the glints occur on the $+z$ surface in this case, both methods can have similar estimation results in the early phase. However, the glint after 50~min, which occurs on the $-y$ surface, causes the single surface UKF to suffer from an estimation error and fail to converge as shown by the blue lines in Fig.~\ref{fig:eulerErrRep01}. On the other hand, the multiple surface IMM provides better estimation accuracy. This result clearly indicates that the wrong hypothesis of the single surface UKF leads to a large estimation error.
Figure~\ref{fig:modeProbRep01} shows the time history of the mode probabilities $w_{j}$ for the multiple surface IMM method. Initially, all probabilities are equal. During glint events, the mode probabilities spike because the likelihood of each filter fluctuates. The mixing step of the IMM ensures that no filter is permanently discarded, allowing adaptation when the glint source transitions between surfaces.


\subsection{Case 2: Glints occur on multiple surfaces}
\begin{figure}[tb]
  \centering
  \includegraphics[width=7.5cm]{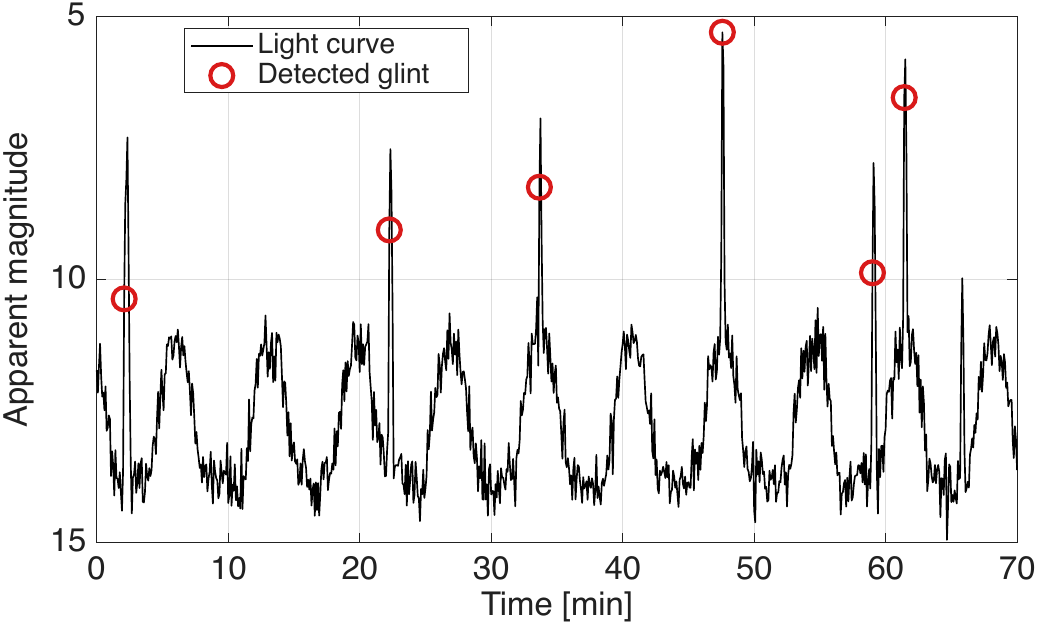}
  \caption{Light curve of the box satellite (Case 2).}
  \label{fig:lightcurveRep02}
\end{figure}

\begin{figure}[tb]
  \centering
  \includegraphics[width=7.5cm]{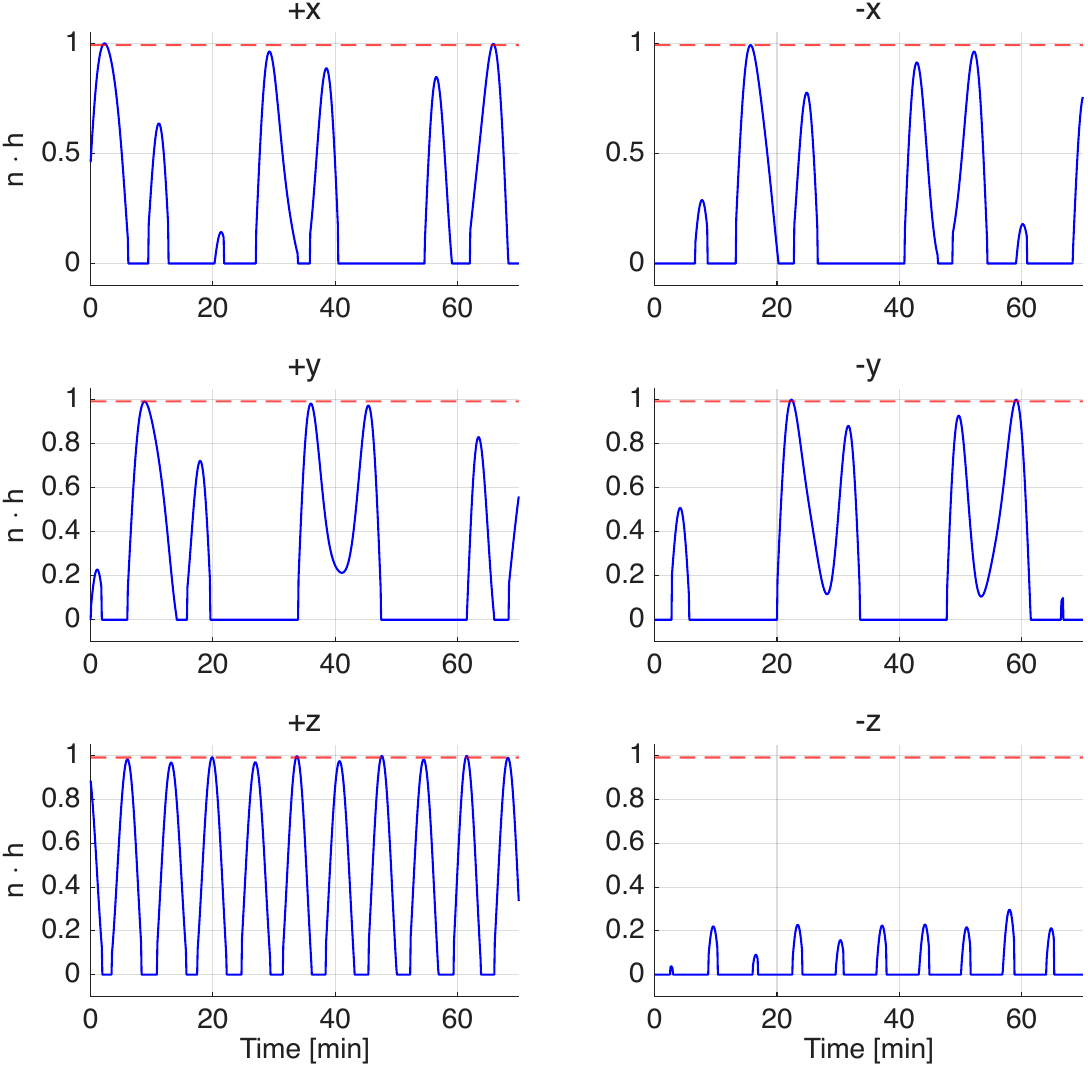}
  \caption{The inner product between each normal vector and the half vector (Case 2).}
  \label{fig:halfVecRep02}
\end{figure}

Figures~\ref{fig:lightcurveRep02} and~\ref{fig:halfVecRep02} show the simulated light curve and the inner product between each normal vector and the half vector for Case 2, respectively.
This is a more challenging case because several glints occur and the glint source surface changes over time as shown in Fig.~\ref{fig:halfVecRep02}.

\begin{figure}[tb]
  \centering
  \includegraphics[width=7.5cm]{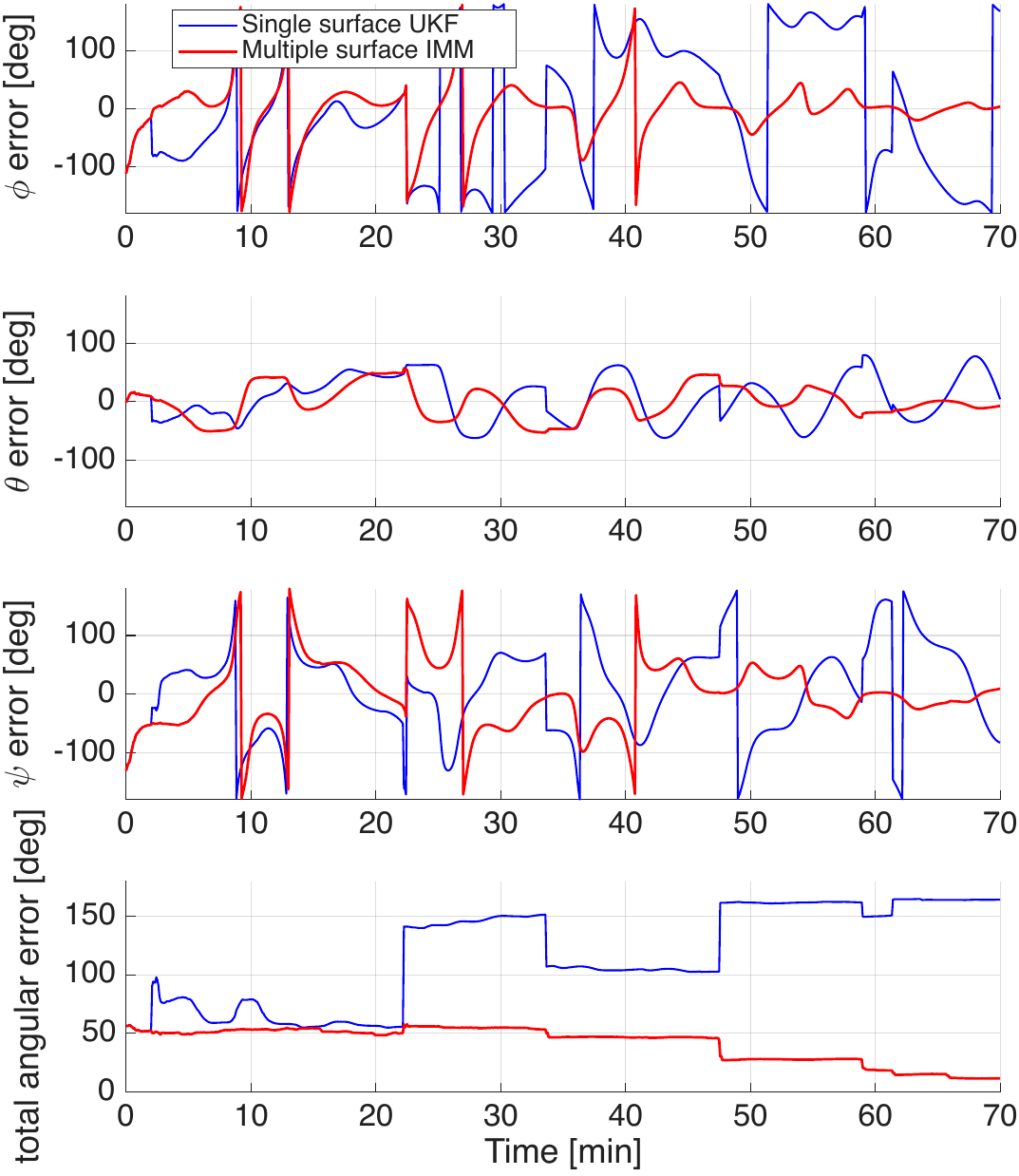}
  \caption{Euler angle error using multi surface IMM and single surface IMM (Case 2).}
  \label{fig:eulerErrRep02}
\end{figure}

\begin{figure}[tb]
  \centering
  \includegraphics[width=7.5cm]{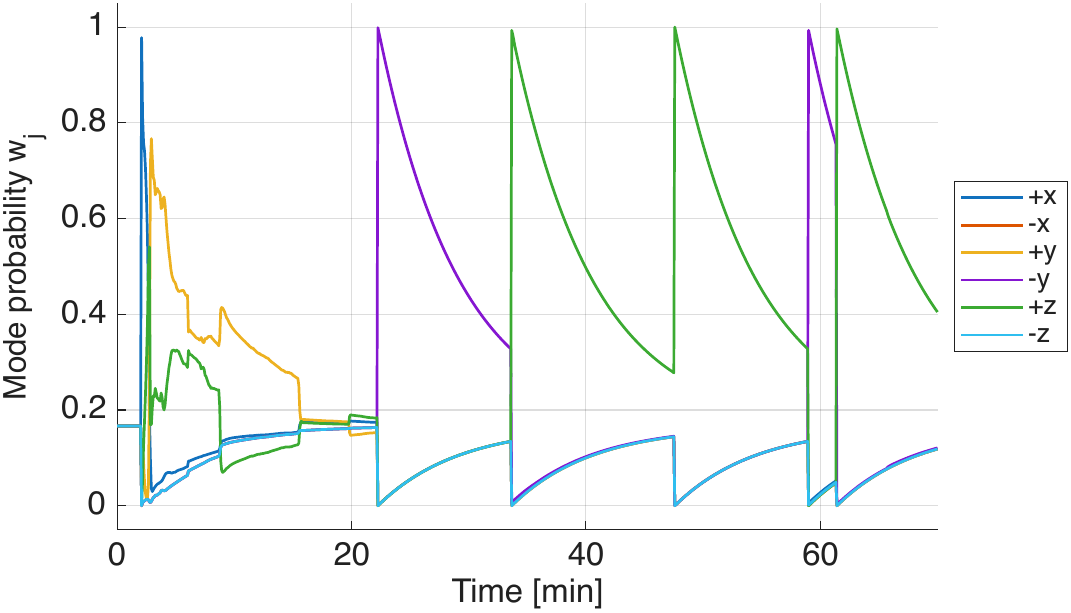}
  \caption{Time history of mode probabilities $w_{j}$ for the multiple surface IMM method (Case 2).}
  \label{fig:modeProbRep02}
\end{figure}

Figure~\ref{fig:eulerErrRep02} shows the attitude estimation errors for the single surface UKF and the multi surface IMM method, including the total attitude error $\Phi$ in the bottom.
Figure~\ref{fig:modeProbRep02} shows the time history of the mode probabilities $w_{j}$ for the multiple surface IMM method, where the mode probabilities at the first glint are not uniquely determined.
As the estimation sequence proceeds, the mode probabilities adapt to the changing glint source, and the total attitude error of the IMM decreases as shown in Fig.~\ref{fig:eulerErrRep02}.
On the other hand, the single surface UKF suffers from a large error and fails to converge, which is expected given the incorrect surface assumption.
The converged residual errors for both cases are summarized in Table~\ref{tb:residual}. The multiple surface IMM attains a total attitude error of $6.4^\circ$ in Case 1 and $11.6^\circ$ in Case 2, whereas the single surface UKF remains above $140^\circ$ in both cases.

\begin{table}[tb]
  \centering
  \caption{Converged residual errors for Cases 1 and 2: mean absolute Euler angle error and total attitude error $\Phi$ (mean $\pm$ standard deviation over the last 5~min of the simulation).}
  \label{tb:residual}
  \begin{tabular}{llcccc}
    \hline\hline
    Case & Method & $|\phi|$ [deg] & $|\theta|$ [deg] & $|\psi|$ [deg] & $\Phi$ [deg] \\
    \hline
    Case 1 & IMM        & $2.1\pm2.5$ & $3.7\pm3.7$  & $5.3\pm5.0$   & $6.4\pm0.3$  \\
           & Single UKF & $152\pm6$   & $13\pm13$    & $15\pm15$     & $148\pm0.2$  \\
    \hline
    Case 2 & IMM        & $6.9\pm5.5$ & $7.1\pm5.4$  & $10.3\pm11.0$ & $11.6\pm0.2$ \\
           & Single UKF & $167\pm164$ & $52\pm53$    & $51\pm46$     & $164\pm0.1$  \\
    \hline\hline
  \end{tabular}
\end{table}

The computational cost of the methods was measured in our MATLAB implementation (MATLAB R2025a) on a desktop computer with an Apple Silicon CPU. For the present scenario, the multiple surface IMM with six parallel filters requires approximately 5~ms per time step, whereas the single surface UKF requires approximately 1~ms per time step; the cost of the IMM scales with the number of parallel filters. Both are well within the 5~s sampling interval, and the additional cost of the IMM does not hinder its application at the considered observation cadence.

\subsection{Sensitivity analysis}
The transition probability $p_{\rm same}$ is the only tuning parameter introduced by the IMM formulation, and an appropriate value is generally not known a priori. The following sensitivity analysis is thus performed to assess how strongly the estimation performance depends on $p_{\rm same}$ and, in particular, whether the proposed method requires careful tuning of this parameter for the considered scenario.
For the 100 Monte Carlo runs, the initial attitude error magnitude is drawn from $U[0,80]$~deg about a uniformly distributed random axis, and the final estimation accuracy is evaluated by the total attitude error $\Phi$ averaged over the last 10 time steps. Because the filter either converges to within a few degrees or settles on an incorrect solution near 150~deg, the outcome is bimodal and the mean is dominated by the divergent tail. The median total attitude error and the convergence rate, defined as the fraction of trials with a final $\Phi$ below 10~deg, are reported.

Table~\ref{tb:sensitivity} summarizes the Monte Carlo results. The single surface UKF converges in only 10\% of the trials, with a median total error of 86~deg, because it cannot resolve which surface causes each glint. In contrast, the proposed IMM converges in 70--73\% of the trials with a median total error of about 4--5~deg for $p_{\rm same}$ between 0.30 and 0.99. When the IMM converges, its mean total error is 3.4~deg, consistent with the representative trials of Cases~1 and 2. The estimation performance is thus insensitive to $p_{\rm same}$ over a wide range, which is a favorable property because it eliminates the need for precise tuning of the transition probability matrix. When $p_{\rm same}=1.00$, however, the IMM reduces to the MMAE and the convergence rate drops to 40\%, which demonstrates the importance of the mixing step. That is, without the interaction among filters, an incorrect surface hypothesis is more likely to cause divergence.

\begin{figure}[tb]
  \centering
  \includegraphics[width=7.5cm]{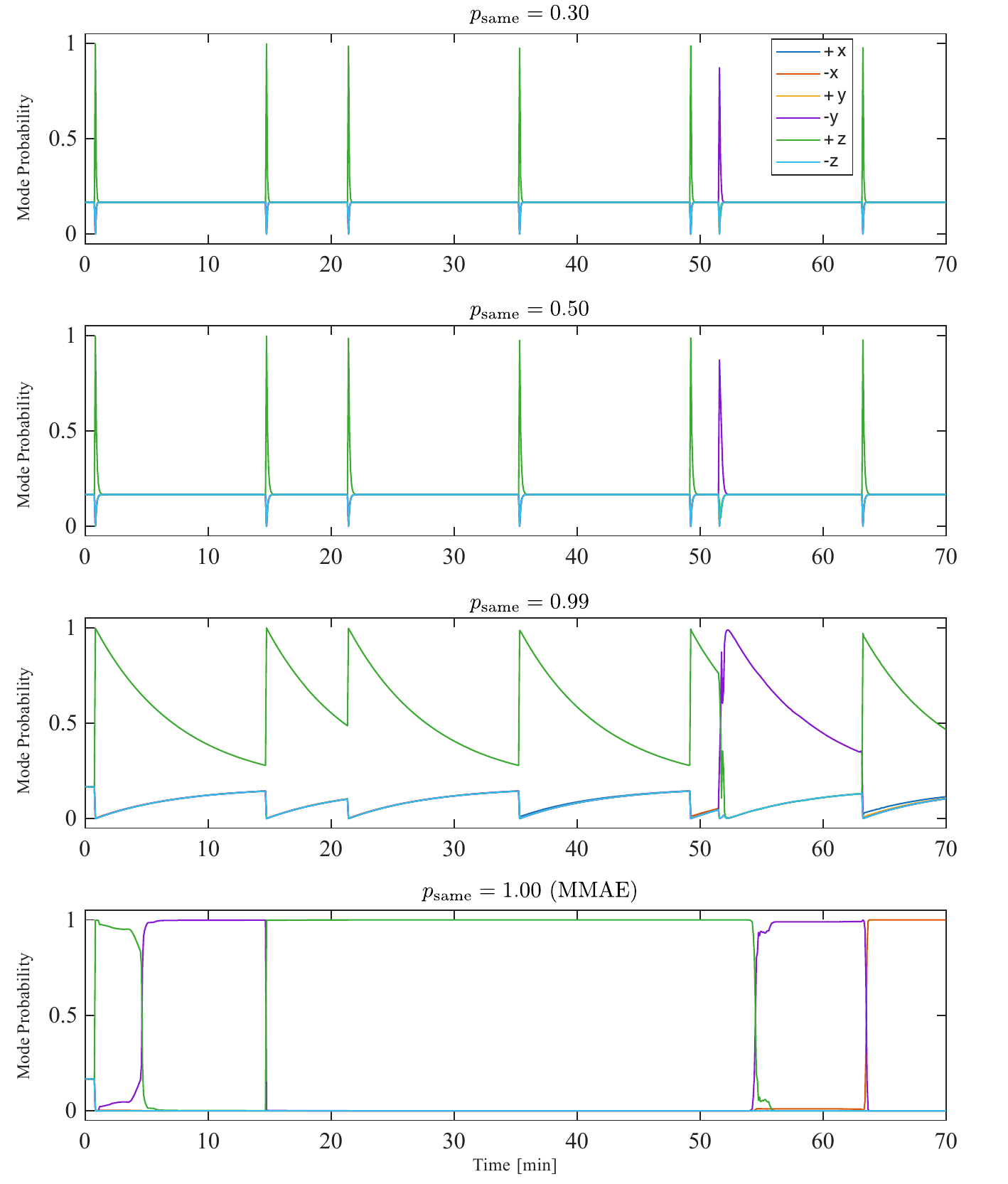}
  \caption{Time history of mode probabilities $w_{j}$ for different values of $p_{\rm same}$.}
  \label{fig:modeProbComparison}
\end{figure}

\begin{figure}[tb]
  \centering
  \includegraphics[width=7.5cm]{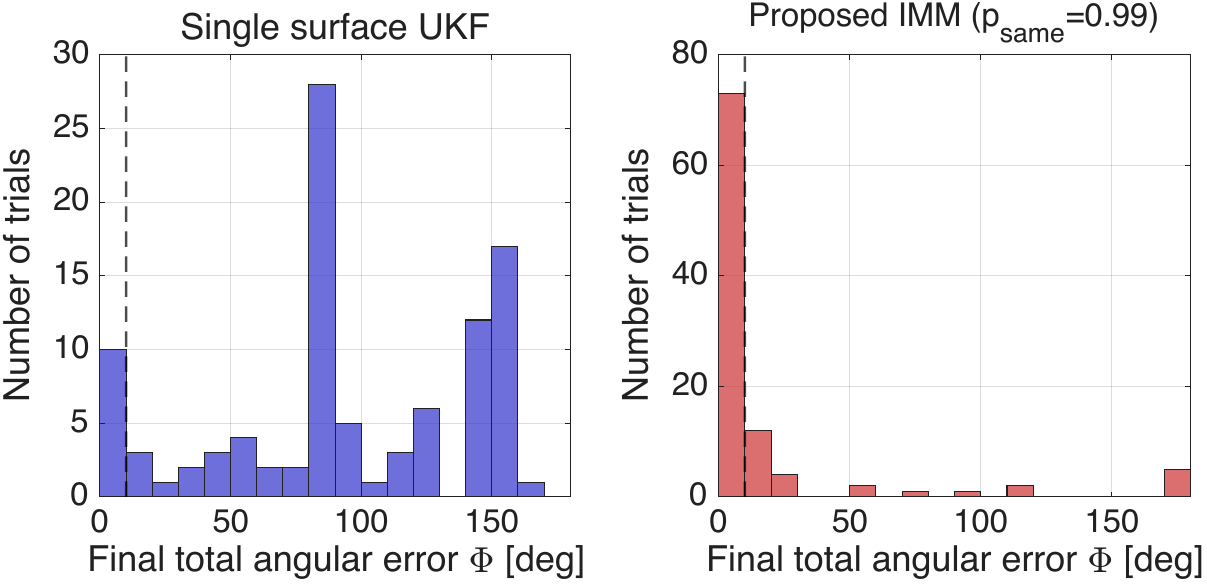}
  \caption{Distribution of the final total attitude error $\Phi$ for the single surface UKF and the proposed IMM method ($p_{\rm same}=0.99$) over the 100 Monte Carlo trials.}
  \label{fig:rmsErrDistMC}
\end{figure}

Figure~\ref{fig:modeProbComparison} shows the time history of the mode probabilities $w_{j}$ for different values of $p_{\rm same}$. Between glint events, the innovation-based likelihood update at each time step drives the mode probabilities toward a uniform distribution regardless of $p_{\rm same}$, so that the mode probabilities are effectively identical at each glint event, which explains the insensitivity to $p_{\rm same}$. Figure~\ref{fig:rmsErrDistMC} shows the distribution of the final total attitude error for the single surface UKF and the proposed IMM method with $p_{\rm same}=0.99$. The errors of the single surface UKF are concentrated at large values, whereas those of the IMM are concentrated near zero.

Table~\ref{tb:initErr} reports the convergence rate of the IMM ($p_{\rm same}=0.99$) as a function of the initial attitude error magnitude, over the deliberately demanding test in which the error is as large as 80~deg. For initial errors below 30~deg, the IMM converges in 97\% of the trials, and even for very large errors of 60--80~deg it still recovers the attitude in 39\% of the trials, whereas the single surface UKF converges in at most 14\% of the trials in any error range. The proposed method thus enlarges the range of initial errors from which the attitude can be recovered.

\begin{table}[tb]
  \centering
  \caption{Total attitude error and convergence rate over 100 Monte Carlo trials for the single surface UKF and the proposed IMM method with different values of $p_{\rm same}$.}
  \label{tb:sensitivity}
  \begin{tabular}{lccc}
    \hline \hline
    Method & $p_{\rm same}$ & Median $\Phi$ [deg] & Conv. rate [\%] \\
    \hline
    Single surface UKF & --- & 86.1 & 10 \\
    \hline
     & 0.30 &  5.2 & 72 \\
    IMM & 0.50 &  4.9 & 70 \\
     & 0.99 &  4.2 & 73 \\
    MMAE & 1.00 & 13.3 & 40 \\
    \hline \hline
  \end{tabular}
\end{table}

\begin{table}[tb]
  \centering
  \caption{Convergence rate of the proposed IMM method ($p_{\rm same}=0.99$) as a function of the initial attitude error magnitude, over the 100 Monte Carlo trials.}
  \label{tb:initErr}
  \begin{tabular}{lccc}
    \hline \hline
    Initial error [deg]   & $[0,30)$ & $[30,60)$ & $[60,80)$ \\
    \hline
    Number of trials      & 39 & 33 & 28 \\
    Convergence rate [\%] & 97 & 73 & 39 \\
    \hline \hline
  \end{tabular}
\end{table}


\subsection{Resilience to a false glint}
In an operational setting, a glint may be detected falsely owing to a sensor anomaly or external noise. Because the detection relies on a sudden change in the apparent magnitude, a spurious brightness change is registered as a glint at an epoch where no surface satisfies the glint geometry, which triggers an erroneous likelihood update and a projection onto an incorrect constraint. To assess the resilience of the proposed method, a false glint is injected into the Case 1 scenario by introducing a spurious 5~mag brightness change at $t=31.9$~min, a quiet epoch at which no surface is aligned with the half vector. The initial estimate error is set to 3~deg so that the response is examined from a well-tracked state.

\begin{figure}[tb]
  \centering
  \includegraphics[width=8.4cm]{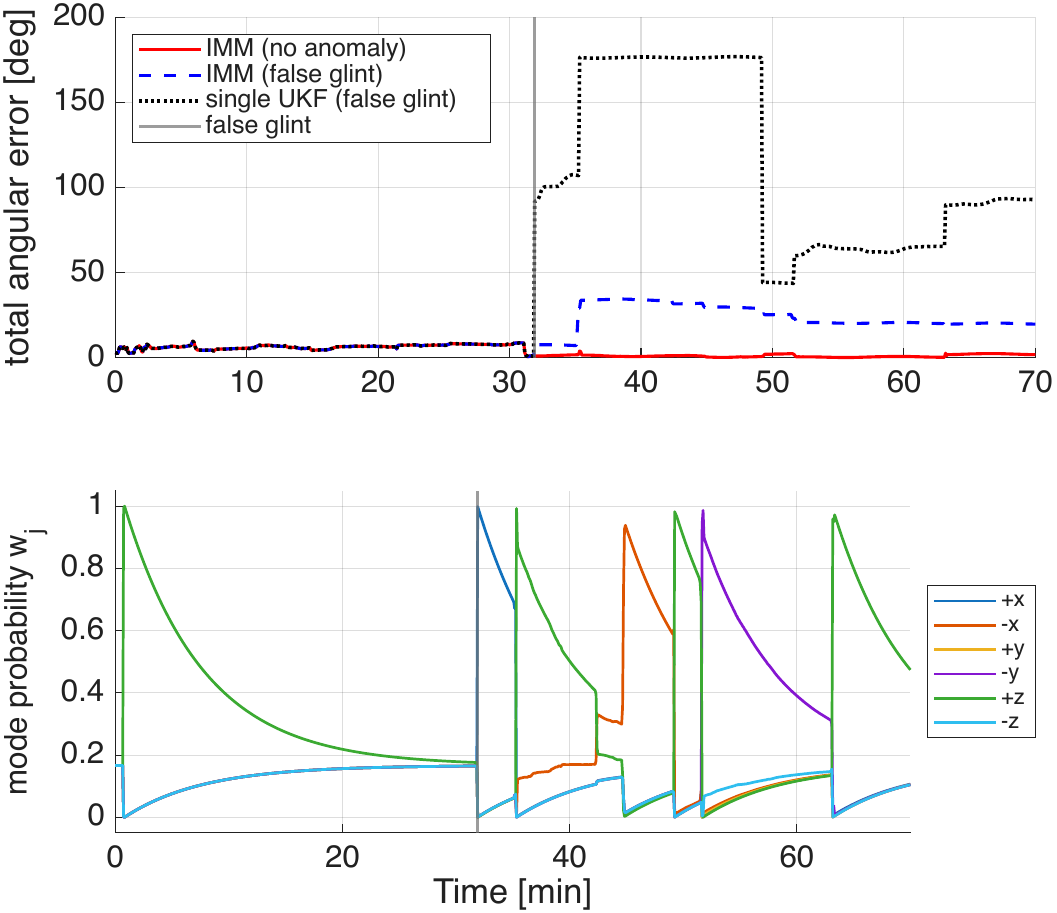}
  \caption{Resilience to a false glint injected at $t=31.9$~min (gray vertical line).}
  \label{fig:falseGlint}
\end{figure}

Figure~\ref{fig:falseGlint} shows the simulation result. At the false glint, the mode probability of the $+x$ filter spikes. However, because no surface is geometrically consistent with a glint, the corresponding likelihood remains small, and the mixing step prevents the incorrect hypothesis from dominating. Consequently, the total attitude error of the IMM remains bounded and is gradually reduced as subsequent genuine glints re-inform the filter. In contrast, the single surface UKF, which has only the $+z$ hypothesis, applies the spurious projection directly and is immediately corrupted, diverging to more than 90~deg. This result demonstrates that the interaction among the filters makes the IMM robust to anomalous observations, whereas a single-hypothesis filter is not.

\section{Conclusions}
This paper proposes an adaptive attitude estimation method for space objects with multiple surfaces using light curves and glint constraints. When a glint is observed from a multiple-surface object, the surface causing the glint is ambiguous, which prevents attitude estimation using the glint constraint. To resolve this ambiguity, this paper employs the interacting multiple model (IMM) algorithm, where each filter assumes a different surface as the glint source. The mixing step allows the filters to interact with each other, enabling adaptation to the time-varying glint source as the object rotates. The likelihood-based mode probability update identifies the correct surface that causes glint and computes the combined attitude estimate as a weighted average.
Numerical simulations for a box satellite in a geosynchronous orbit demonstrate that the proposed IMM method improves the attitude estimation. Over 100 Monte Carlo runs with initial errors of up to 80~deg, the IMM converges in 73\% of the trials with a median total attitude error of 4.2~deg, compared with 10\% for the single surface UKF, and the mixing step is shown to be essential, as the convergence rate drops to 40\% when it is removed (the MMAE). The convergence rate depends on the initial error magnitude, remaining at 97\% for errors below 30~deg and at 39\% even for errors as large as 60--80~deg, where the single surface UKF converges in at most 14\%. The IMM thus enlarges the range of initial errors from which the attitude can be recovered. The mode probabilities correctly shift toward the filter corresponding to the true glint-causing surface, validating the effectiveness of the likelihood-based selection and the model interaction. The method is also shown to be resilient to a falsely detected glint. That is, the mixing step keeps the impact of the anomalous observation bounded, whereas a single-hypothesis filter diverges.

\section*{Declaration of generative AI and AI-assisted technologies in the manuscript preparation process}
During the preparation of this work the authors used chatGPT in order to correct typos and improve English grammar. After using this tool, the authors reviewed and edited the content as needed and take full responsibility for the content of the published article.


\bibliographystyle{cas-model2-names}
\bibliography{multiGlint}

\end{document}